\documentclass{article}
\usepackage{spconf,amsmath,graphicx}

\usepackage{graphicx,cite,amssymb,amsmath,psfrag,subfigure}
\usepackage{graphicx}
\usepackage{amssymb}
\usepackage{amsmath}
\usepackage{cite}
\usepackage{subfigure}
\usepackage{mathrsfs}
\usepackage[displaymath,mathlines]{lineno}
\usepackage{color}
\usepackage{tabulary}
\usepackage{multirow}

\newtheorem{algorithm}{Algorithm}

\DeclareMathOperator{\E}{\mathsf{E}}

\newcommand{\EX}[1]{\E\left\{{#1}\right\}}

\newcommand{\CG}[2]{\mathcal{CN}\left({#1},{#2}\right)}

\newcommand{\B}[1]{{\mathbf{#1}}}

\title{Blind Estimation of Effective Downlink Channel Gains in Massive MIMO}
\name{Hien Quoc Ngo \qquad Erik G. Larsson \thanks{This work was
    supported in part by the Swedish Research Council (VR) and
    ELLIIT.}}

\address{Department of Electrical Engineering (ISY), Link\"{o}ping University, 581 83 Link\"{o}ping, Sweden}

\begin{document}
\ninept
\maketitle
\begin{abstract}
We consider the massive MIMO downlink with time-division duplex (TDD)
operation and conjugate beamforming transmission. To reliably decode
the desired signals, the users need to know the effective channel
gain. In this paper, we propose a blind channel estimation method
which can be applied at the users and which does not require any
downlink pilots.  We show that our proposed scheme can substantially
outperform the case where each user has only statistical channel
knowledge, and that the difference in performance is particularly
large in certain types of channel, most notably keyhole
channels. Compared to schemes that rely on downlink pilots (e.g.,
\cite{NLM:13:ACCCC}), our proposed scheme yields more accurate channel
estimates for a wide range of signal-to-noise ratios and avoid
spending time-frequency resources on pilots.
\end{abstract}
\section{Introduction}
\label{sec:intro}

Massive multiple-input multiple-output (MIMO) is one of the most
promising technologies to meet the demands for high throughput and
communication reliability of next generation cellular networks
\cite{Eri:13:MCOM,ZJWZM:14:SSP,LL:14:SP,NZSL:14:SP}. In massive MIMO,
time-division duplex (TDD) operation is preferable since then the
pilot overhead does not depend on the number of base station
antennas. With TDD, the channels are estimated at the base station
through the uplink training. For the downlink, under the assumption of
channel reciprocity, the channels estimated at the base station are
used to precode the data, and the precoded data are sent to the users.
To coherently decode the transmitted signals, each user should have
channel state information (CSI), that is, know its effective channel
from the base station.

In most previous works, the users are assumed to have statistical
knowledge of the effective downlink channels, that is, they know the
mean of the effective channel gain and use this for the signal
detection \cite{YM:13:JSAC,JAMV:11:WCOM}. In these papers, Rayleigh
fading channels were assumed. Under the Rayleigh fading, the effective
channel gains become nearly deterministic (the channel ``hardens'')
when the number of base station antennas grows large, and hence, using
the mean of the effective channel gain for signal detection works very
well. However, in practice, propagation scenarios may be encountered
where the channel does not harden.  In that case, using the mean
effective channel gain may not be accurate enough, and a better
estimate of the effective channel should be used. In
\cite{NLM:13:ACCCC}, we proposed a scheme where the base station (in
addition to the beamformed data) also sent a beamformed downlink pilot
sequence to the users.  With this scheme, a performance improvement
(compared to the case when the \emph{mean} of the effective channel
gain is used) was obtained. However, this scheme requires
time-frequency resources in order to send the downlink pilots. The
associated overhead is proportional to the number of users which can
be in the order of several tens, and hence, in a high-mobility
environment (where the channel coherence interval is short) the
spectral efficiency is significantly reduced.

\textit{Contribution:} In this paper, we consider the massive MIMO
downlink with conjugate beamforming.\footnote{We consider conjugate
  beamforming since it is simple and nearly optimal in many massive
  MIMO scenarios. More importantly, conjugate beamforming can be
  implemented in a distributed manner.}  We propose a scheme with
which the users \emph{blindly} estimate the effective channel gain
from the received data. The scheme exploits the asymptotic properties
of the mean of the received signal power when the number of base
station antennas is large. The accuracy of our proposed scheme is
investigated for two specific, very different, types of channels: (i)
independent Rayleigh fading and (ii) keyhole channels. We show that
when the number of base station antennas goes to infinity, the channel
estimate provided by our scheme becomes exact. Also, numerical results
quantitatively show the benefits of our proposed scheme, especially in
keyhole channels, compared to the case where the mean of the effective
channel gain is used as if it were the true channel gain, and compared
to the case where the beamforming training scheme of
\cite{NLM:13:ACCCC} is used.

\textit{Notation:} We use boldface upper- and lower-case letters to
denote matrices and column vectors, respectively. The superscripts
$()^T$ and $()^H$ stand for the transpose and conjugate transpose,
respectively. The Euclidean norm, the trace, and the expectation
operators are denoted by $\|\cdot\|$, $\text{Tr}\left(\cdot\right)$,
and $\EX{\cdot}$, respectively. The notation $\mathop \to \limits^{P}$
means convergence in probability, and $\mathop \to \limits^{a.s.}$
means almost sure convergence. Finally, we use $z \sim
\CG{0}{\sigma^2}$ to denote a circularly symmetric complex Gaussian
random variable (RV) $z$ with zero mean and variance $\sigma^2$.

\section{System Model}

Consider the downlink of a massive MIMO system. An $M$-antenna base
station serves $K$ single-antenna users, where $M\gg K\gg 1$. The base
station uses conjugate beamforming to simultaneously transmit data to
all $K$ users in the same time-frequency resource. Since we focus on
the downlink channel estimation here, we assume that the base station
perfectly estimates the channels in the uplink training phase. (In
future work, this assumption may be relaxed.)  Denote by $\B{g}_k$ the
$M\times 1$ channel vector between the base station and the $k$th
user. The channel $\B{g}_k$ results from a combination of small-scale
fading and large-scale fading, and is modeled as:
\begin{align}\label{eq:sys1}
\B{g}_k = \sqrt{\beta_k}\B{h}_k,
\end{align}
where $\beta_k$ represents large-scale fading which is constant over
many coherence intervals, and $\B{h}_k$ is an $M\times 1$ small-scale
channel vector. We assume that the elements of $\B{h}_k$ are
i.i.d.\ with zero mean and unit variance.

Let $s_k$, $\EX{|s_k|^2}=1$, $k=1, \ldots, K$, be the symbol
intended for the $k$th user. With conjugate beamforming, the
$M\times 1$ precoded signal vector is given by
\begin{align}\label{eq:sys2}
\B{x} = \sqrt{\alpha}\B{G}\B{s},
\end{align}
where $\B{s}\triangleq [s_1, s_2, \ldots, s_K]^T$,
$\B{G}\triangleq [\B{g}_1 \ldots \B{g}_K]$ is an $M\times K$
channel matrix between the $K$ users and the base station, and
$\alpha$ is a normalization constant chosen to satisfy the average
power constraint at the base station: $$\EX{\|\B{x}\|^2}=\rho.$$
Hence,
\begin{align}\label{eq:alp2}
\alpha = \frac{\rho}{\EX{\mathrm{Tr}\left(\B{G}\B{G}^H \right)}}.
\end{align}
The signal received at the $k$th user is
\begin{align}\label{eq:sys3}
y_k &= \B{g}_k^H\B{x} + n_k = \sqrt{\alpha}\B{g}_k^H\B{G}\B{s} + n_k\nonumber\\
    &=  \sqrt{\alpha}\|\B{g}_k\|^2s_k + \sqrt{\alpha}\sum_{k'\neq k}^K \B{g}_k^H\B{g}_{k'}s_{k'} +
    n_k,
\end{align}
where $n_k\sim\CG{0}{1}$ is the additive Gaussian noise at the
$k$th user. Then, the desired signal $s_k$ is decoded.

\section{Proposed Downlink Blind Channel Estimation Technique}

The $k$th user wants to detect $s_k$ from $y_k$ in
\eqref{eq:sys3}. For this purpose, it needs to know the effective channel gain
$\|\B{g}_k\|^2$. If the channel is Rayleigh fading, then by the
law of large numbers, we have $$\frac{1}{M}\|\B{g}_k\|^2 \mathop
\to\limits^{P} \beta_k,$$ as $M\to\infty$. This implies that when
$M$ is large, $\|\B{g}_k\|^2 \approx M\beta_k$ (we say that the channel
\emph{hardens}). So we can use the statistical properties of the
channel, i.e., use $\EX{\|\B{g}_k\|^2}=M\beta_k$ as a good estimate of
$\|\B{g}_k\|^2$ when detecting $s_k$. This assumption is widely made in
the massive MIMO literature. However, in practice, the channel
is not always Rayleigh fading, and does not always harden when $M\to\infty$. For example,
consider a keyhole channel, where the small-scale fading component $\B{h}_k$ is modeled as follows
\cite{SL:03:IT,ZJWM:11:WCOM}:
\begin{align}\label{eq:keyhole1}
\B{h}_k &= \nu_k\bar{\B{h}}_k,
\end{align}
where $\nu_k$ and the $M$ elements of $\bar{\B{h}}_k$ are i.i.d.\
$\CG{0}{1}$ RVs. For the keyhole channel \eqref{eq:keyhole1}, by
 the law of large numbers, we have
$$\frac{1}{M}\|\B{g}_k\|^2 - \beta_k|\nu_k|^2 \mathop \to\limits^{P} 0,$$
which is not deterministic, and hence the channel does not harden.
In this case, using $\EX{\|\B{g}_k\|^2}=M\beta_k$ as an estimate of the true
effective channel $\|\B{g}_k\|^2$ to detect $s_k$ may result in poor performance.

For the reasons explained, it is desirable that the users estimate
their effective channels.  One way to do this is to have the base
station transmit beamformed downlink pilots as proposed in
\cite{NLM:13:ACCCC}. With this scheme, at least $K$ downlink pilot
symbols are required.  This can significantly reduce the spectral
efficiency. For example, suppose $M=300$ antennas serve $K=50$
terminals, in a coherence interval of length $200$ symbols. If half of
the coherence interval is used for the downlink, then with the
downlink beamforming training of \cite{NLM:13:ACCCC}, we need to spend
at least $50$ symbols for sending pilots. As a result, less than $50$
of the $100$ downlink symbols are used for payload in each coherence
interval, and the insertion of the downlink pilots reduces the
overall (uplink+downlink) spectral efficiency by a factor of $1/4$.

In what follows, we propose a blind channel estimation method which
does not require any downlink pilots.

\subsection{Mathematical Preliminaries} \label{subsec: MP}

Consider the average  power of the received signal at the $k$th user (averaged over $\B{s}$ and $n_k$). From \eqref{eq:sys3}, we have
\begin{align}\label{eq:po1}
\EX{|y_k|^2}     &=  \alpha\|\B{g}_k\|^4 + \alpha\sum_{k'\neq k}^K \left|\B{g}_k^H\B{g}_{k'}\right|^2 + 1.
\end{align}
The second term of \eqref{eq:po1} can be rewritten as
\begin{align}\label{eq:po2}
\alpha\sum_{k'\neq k}^K \left|\B{g}_k^H\B{g}_{k'}\right|^2
    =
    \alpha\sum_{k'\neq k}^K \B{g}_{k'}^H\B{g}_{k}\B{g}_{k}^H\B{g}_{k'}=
    \alpha\tilde{\B{g}}_k^H\B{A}\tilde{\B{g}}_k,
\end{align}
where $\tilde{\B{g}}_k\triangleq [\B{g}_1^T ~ \ldots ~
\B{g}_{k-1}^T ~ \B{g}_{k+1}^T ~ \ldots ~ \B{g}_K^T]^T$, and
$\B{A}$ is an $M(K-1)\times M(K-1)$ block-diagonal matrix whose
$(i,i)$-block is the $M\times M$ matrix $\B{g}_k\B{g}_k^H$. Since
$\B{A}$ and $\tilde{\B{g}}_k$ are independent, as
$M(K-1)\to\infty$, the Trace lemma gives \cite{WCDS:12:IT}
\begin{align}\label{eq:po3}
\frac{1}{M(K-1)}\sum_{k'\neq k}^K \left|\B{g}_k^H\B{g}_{k'}\right|^2
    -
    \frac{1}{M(K-1)} \sum_{k'\neq k}^K\beta_{k'}\|\B{g}_k\|^2 ~ ~ \mathop \to \limits^{a.s.} ~ 0.
\end{align}
Substituting \eqref{eq:po3} into \eqref{eq:po1}, as $M(K-1)\to\infty$, we have
\begin{align}\label{eq:po1a}
\frac{\EX{|y_k|^2}}{M(K-1)}  &-\frac{1}{M(K-1)} \left( \alpha\|\B{g}_k\|^4 + \alpha\sum_{k'\neq k}^K\beta_{k'}\|\B{g}_k\|^2 + 1\right) \nonumber\\&\hspace{5.5cm}\mathop \to\limits^{a.s.} 0.
\end{align}
The above result implies that when $M$ and $K$ are large,
\begin{align}\label{eq:po4}
\EX{|y_k|^2}  \approx \alpha\|\B{g}_k\|^4 + \alpha\sum_{k'\neq k}^K\beta_{k'}\|\B{g}_k\|^2 + 1.
\end{align}
Therefore, the effective channel gain
$\|\B{g}_k\|^2$ can be estimated from $\EX{|y_k|^2}$ by solving
the quadratic equation \eqref{eq:po4}.

\subsection{Downlink Blind Channel Estimation Algorithm}

As discussed in Section~\ref{subsec: MP}, we can estimate the
effective channel gain $\|\B{g}_k\|^2$ by solving the quadratic
equation \eqref{eq:po4}. It is then required that the $k$th user knows
$\alpha$, $\sum_{k'\neq k}^K\beta_{k'}$, and $\EX{|y_k|^2}$. We assume
that the $k$th user knows $\alpha$ and $\sum_{k'\neq
  k}^K\beta_{k'}$. This assumption is reasonable since the terms
$\alpha$ and $\sum_{k'\neq k}^K\beta_{k'}$ depend on the large-scale
fading coefficients, which stay constant over many coherence
intervals. The $k$th user can estimate these terms, or the base
station may inform the $k$th user about them. Regarding
$\EX{|y_k|^2}$, in practice, it is unavailable.  However, we can use
the received samples during a whole coherence interval to form a sample estimate of
$\EX{|y_k|^2}$ as follows:
\begin{align}\label{eq:al1}
\EX{|y_k|^2}  \approx \xi_k \triangleq \frac{|y_k(1)|^2+ |y_k(2)|^2+ \ldots+ |y_k(T)|^2}{T},
\end{align}
where $y_k(n)$ is the $n$th receive sample, and $T$ is the length
(in symbols) of the coherence interval used for the downlink
transmission.

The algorithm for estimating $\|\B{g}_k\|^2$ is summarized as follows:
\begin{algorithm} (Proposed blind downlink  channel estimation
method)
\begin{description}
  \item[1.] Using a data block of $T$ samples, compute $\xi_k$ as \eqref{eq:al1}.
  \item[2.] The channel estimate of $\|\B{g}_k\|^2$, denoted by $a_k$, is determined as
\begin{align}\label{eq:alg1}
\hspace{-0.5cm}a_k \!=\! \frac{-\alpha\sum_{k'\neq k}^K\beta_{k'}
\!+\! \sqrt{\alpha^2\!\left(\!\sum_{k'\neq k}^K\beta_{k'}
\!\right)^2 \!+\! 4\alpha(\xi_k\!-\!1)}}{2\alpha}.
\end{align}
\end{description}
\end{algorithm}
Note that $a_k$ in \eqref{eq:alg1} is the positive root of the
quadratic equation: $\xi_k=\alpha a_k^2 + \alpha\sum_{k'\neq
k}^K\beta_{k'}a_k + 1$ which comes from \eqref{eq:po4} and
\eqref{eq:al1}.

\begin{figure}[t]
\begin{minipage}[b]{1.0\linewidth}
  \centering
  \centerline{\includegraphics[width=8.0 cm]{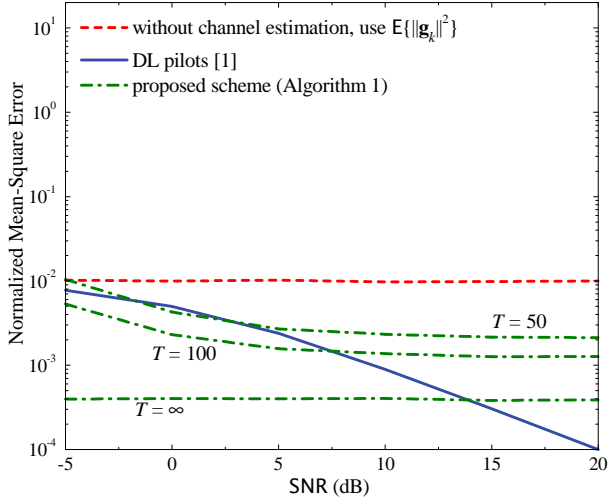}}
\end{minipage}
\caption{Normalized MSE versus $\mathsf{SNR}$ for different
channel estimation schemes, for Rayleigh fading channels.}
\label{fig:1}
\end{figure}

\subsection{Asymptotic Performance Analysis} \label{Sec:PA}

In this section, we analyze the accuracy of our proposed scheme for two specific propagation environments: Rayleigh
fading and keyhole channels. For keyhole channels, we use the model
\eqref{eq:keyhole1}. We assume that the $k$th user perfectly estimates
$\EX{|y_k|^2}$. This is true when the number of symbols of the coherence interval allocated to the
downlink, $T$, is large. In the numerical results, we
shall show that the estimate of $\EX{|y_k|^2}$ in \eqref{eq:al1} is
very close to $\EX{|y_k|^2}$ even for modest values of $T$ (e.g.\
$T\approx 100$ symbols). With the assumption $\xi_k=\EX{|y_k|^2}$,
from \eqref{eq:po1} and \eqref{eq:alg1}, the estimate of
$\|\B{g}_k\|^2$ can be written as:
\begin{align}\label{eq:PE1}
a_k = -\frac{\sum_{k'\neq
k}^K\beta_{k'}}{2}+\sqrt{\left(\frac{\sum_{k'\neq
k}^K\beta_{k'}}{2}+\|\B{g}_k\|^2\right)^2 + \epsilon_k},
\end{align}
where
\begin{align}\label{eq:PE2}
\epsilon_k \triangleq \sum_{k'\neq k}^K \left|\B{g}_k^H\B{g}_{k'}\right|^2 - \left(\sum_{k'\neq k}^K\beta_{k'} \right) \|\B{g}_k\|^2.
\end{align}
We can see from \eqref{eq:PE1} that if
$|\epsilon_k|\ll\left(\frac{\sum_{k'\neq
k}^K\beta_{k'}}{2}+\|\B{g}_k\|^2\right)^2$, then $a_k\approx
\|\B{g}_k\|^2$. In order to see under what conditions $|\epsilon_k|\ll\left(\frac{\sum_{k'\neq
k}^K\beta_{k'}}{2}+\|\B{g}_k\|^2\right)^2$, we consider $\varrho_k$ which is defined as:
\begin{align}\label{eq:PE2b}
\varrho_k \triangleq \EX{\left|\epsilon_k /\EX{\left(\frac{1}{2}\sum_{k'\neq k}^K\beta_{k'}+\|\B{g}_k\|^2\right)^2} \right|^2}.
\end{align}
Hence,
\begin{align}\label{eq:PE3}
\varrho_k \!\!=\!\left\{\!\!
             \begin{array}{l}
             \!\! \frac{M(M+1)\beta_k^2\sum\limits_{k'\neq k}^K\beta_{k'}^2}{\left(\frac{1}{4}\bar{\beta}_k^2 + M\beta_{k}\!\!\!\sum\limits_{k'=1}^K\!\!\beta_{k'} + \beta_k^2M^2\right)^2}, ~ \text{for Rayleigh fading channels}, \\
              \!\! \frac{6M(M+1)\beta_k^2\sum\limits_{k'\neq k}^K\beta_{k'}^2}{\left(\frac{1}{4}\bar{\beta}_k^2 + M\beta_{k}\sum\limits_{k'=1}^K\beta_{k'} + \beta_k^2M(2M+1)\right)^2}, ~ \text{for keyhole channels}, \\
             \end{array}
           \right.
\end{align}
where $\bar{\beta}_k\triangleq\sum_{k'\neq k}^K\beta_{k'}$. The
detailed derivations of \eqref{eq:PE3} are presented in the Appendix. We
can see that $\varrho_k=  O(1/M^2)$. Thus, when
$M\gg 1$, $|\epsilon_k|$ is much smaller than
$\left(\frac{\sum_{k'\neq
k}^K\beta_{k'}}{2}+\|\B{g}_k\|^2\right)^2$. As a result, our
proposed channel estimation scheme is expected to work well.

%

\begin{figure}[t]
\begin{minipage}[b]{1.0\linewidth}
  \centering
  \centerline{\includegraphics[width=8.0 cm]{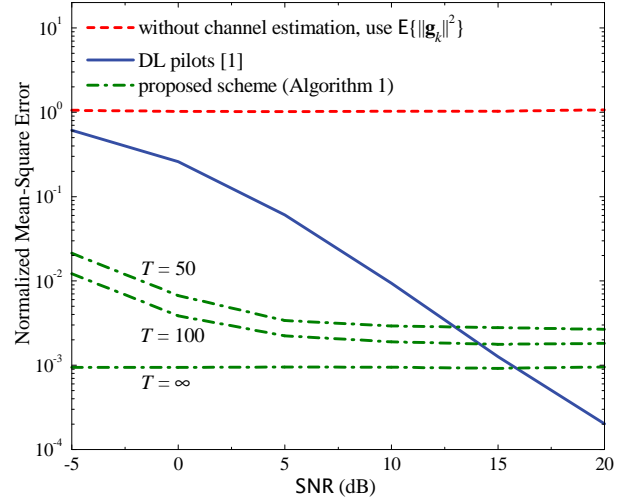}}
\end{minipage}
\caption{Normalized MSE versus $\mathsf{SNR}$ for different
channel estimation schemes, for keyhole channels.} \label{fig:2}
\end{figure}

\section{Numerical Results}

In this section, we provide  numerical results to evaluate our
proposed channel estimation scheme for finite $M$. As performance metric we consider the normalized
mean-square error (MSE) at the $k$th user:
\begin{align}\label{eq:Nr1}
{\tt MSE}_k \triangleq \EX{\left|\frac{a_k-\|\B{g}_k\|^2}{\EX{\|\B{g}_k\|^2}}\right|^2}.
\end{align}
For the simulation, we choose $M=100$, $K=20$, and $\beta_k=1, \forall
k =1, \ldots, K$. We define $\mathsf{SNR}\triangleq \rho$.
Figures~\ref{fig:1} and \ref{fig:2} show the normalized MSE versus
$\mathsf{SNR}$ for Rayleigh fading and keyhole channels,
respectively. The curves labeled ``without channel estimation, use
$\EX{\|\B{g}_k\|^2}$'' represent the case when the $k$th user uses the
statistical properties of the channels, i.e., it uses
$\EX{\|\B{g}_k\|^2}$ as estimate of $\|\B{g}_k\|^2$.  The curves ``DL
pilots \cite{NLM:13:ACCCC}'' represent the case when the beamforming
training scheme of \cite{NLM:13:ACCCC} with MMSE channel estimation is
applied.  The curves ``proposed scheme (Algorithm~1)'' represent our
proposed scheme for different $T$ ($T=\infty$ implies that the $k$th
user perfectly knows $\EX{|y_k|^2}$).
For the beamforming training scheme, the duration of the downlink
training is $K$. For our proposed blind channel estimation scheme,
$s_k, k=1, \ldots, K$, are random 4-QAM symbols.

We can see that in Rayleigh fading channels, the MSEs of the three
schemes are comparable.  Using $\EX{\|\B{g}_k\|^2}$ in lieu of the
true ${\|\B{g}_k\|^2}$ for signal detection works rather
well. However, in keyhole channels, since the channels do not harden,
the MSE when using $\EX{\|\B{g}_k\|^2}$ as estimate of
${\|\B{g}_k\|^2}$ is very large. In both propagation environments, our
proposed scheme works very well. For a wide range of SNRs, our scheme outperforms the beamforming training scheme,
even for short coherence intervals  (e.g., $T=100$ symbols). Note again that, with
the beamforming training scheme of \cite{NLM:13:ACCCC}, we additionally have to spend  at
least $K$ symbols on training pilots (this is not accounted for here, since we only evaluated MSE).
By contrast, our proposed scheme does
not requires any resources for downlink training.

\section{Concluding Remarks} \label{Sec:Conclusion}

Massive MIMO systems may encounter propagation conditions when the
channels do not harden.  Then, to facilitate detection of the data in
the downlink, the users need to estimate their effective channel gain
rather than relying on knowledge of the \emph{average} effective
channel gain.  We proposed a channel estimation approach by which the
users can blindly estimate the effective channel gain from the data
received during a coherence interval. The approach is computationally
easy, it does not requires any resource for downlink pilots, it can be
applied regardless of the type of propagation channel, and it performs
very well.

Future work may include studying rate expressions rather than channel
estimation MSE, and taking into account the channel estimation errors
in the uplink. (We hypothesize, that the latter will not qualitatively
affect our results or conclusions.) Blind estimation of $\beta_k$ by
the users may also be addressed.


\section{Appendix}
Here, we provide the proof  of \eqref{eq:PE3}. From \eqref{eq:PE2b}, we have
\begin{align}\label{eq:proof1}
\varrho_k =\EX{\left|\epsilon_k\right|^2} /\EX{\left(\frac{1}{2}\sum_{k'\neq k}^K\beta_{k'}+\|\B{g}_k\|^2\right)^2}^2.
\end{align}
\begin{itemize}
  \item \emph{Rayleigh Fading Channels:}
\end{itemize}
For Rayleigh fading channels, we have
\begin{align}\label{eq:proofray1}
&\EX{\left(\frac{1}{2}\sum_{k'\neq k}^K\beta_{k'}+\|\B{g}_k\|^2\right)^2} = \frac{1}{4}\left(\sum_{k'\neq k}^K\beta_{k'} \right)^2 \nonumber\\&+ \left(\sum_{k'\neq k}^K\beta_{k'} \right)\EX{\|\B{g}_k\|^2} + \EX{\|\B{g}_k\|^4}\nonumber\\
&=\frac{1}{4}\left(\sum_{k'\neq k}^K\beta_{k'} \right)^2 + M\beta_{k}\sum_{k'=1}^K\beta_{k'} + \beta_k^2M^2,
\end{align}
where the last equality follows \cite[Lemma~2.9]{TV:04:FTCIT}. We next compute $\EX{|\epsilon_k|^2}$. From \eqref{eq:PE2}, we have
\begin{align}\label{eq:proofray2}
\EX{|\epsilon_k|^2} &= \EX{\left(\sum_{k'\neq k}^K \left|\B{g}_k^H\B{g}_{k'}\right|^2 \right)^2} + \left(\sum_{k'\neq k}^K\beta_{k'} \right)^2\EX{\|\B{g}_k\|^4} \nonumber\\&- 2\left(\sum_{k'\neq k}^K\beta_{k'} \right)\EX{\sum_{k'\neq k}^K \left|\B{g}_k^H\B{g}_{k'}\right|^2  \|\B{g}_k\|^2}.
\end{align}
We have,
\begin{align}\label{eq:proofray3}
\EX{\left(\sum_{k'\neq k}^K \left|\B{g}_k^H\B{g}_{k'}\right|^2 \right)^2}
\!=\!\EX{\|\B{g}_k\|^4 \left(\sum_{k'\neq k}^K |z_{k'}|^2 \right)^2},
\end{align}
where $z_{k'}\triangleq
\frac{\B{g}_k^H\B{g}_{k'}}{\|\B{g}_k\|}$. Conditioned on $\B{g}_k$,
$z_{k'}$ is complex Gaussian distributed with zero mean and variance
$\beta_{k'}$ which is independent of $\B{g}_k$. Thus,
$z_{k'}\sim\CG{0}{\beta_{k'}}$ and is independent of $\B{g}_k$. This
yields
\begin{align}\label{eq:proofray4}
&\EX{\left(\sum_{k'\neq k}^K \left|\B{g}_k^H\B{g}_{k'}\right|^2 \right)^2}
=\EX{\|\B{g}_k\|^4}\EX{ \left(\sum_{k'\neq k}^K |z_{k'}|^2 \right)^2}\nonumber\\
&=\beta_k^2M\left(M+1\right)\left(\sum_{i\neq k}^K\beta_{i}^2 + \sum_{i\neq k}^K\sum_{j\neq k}^K\beta_i\beta_j\right).
\end{align}
Similarly,
\begin{align}\label{eq:proofray5}
\EX{\sum_{k'\neq k}^K \left|\B{g}_k^H\B{g}_{k'}\right|^2  \|\B{g}_k\|^2}
&=\EX{\|\B{g}_k\|^4}\EX{\sum_{k'\neq k}^K |z_{k'}|^2}\nonumber\\
&=\beta_k^2M\left(M+1\right)\sum_{k'\neq k}^K\beta_{k'}^2.
\end{align}
Substituting \eqref{eq:proofray4}, \eqref{eq:proofray5}, and $\EX{\|\B{g}_k\|^4} =\beta_k^2M(M+1)$ into \eqref{eq:proofray2}, we obtain
\begin{align}\label{eq:proofray6}
\EX{|\epsilon_k|^2} &= M(M+1)\beta_k^2\sum_{k'\neq k}^K\beta_{k'}^2.
\end{align}
Inserting \eqref{eq:proofray1} and \eqref{eq:proofray6} into \eqref{eq:proof1}, we obtain \eqref{eq:PE3} for the Rayleigh fading case.
\begin{itemize}
  \item \emph{Keyhole Channels:}
\end{itemize}
By using the fact that
\begin{align}\label{eq:proofkh1}
z_{k'} = \frac{\B{g}_k^H\B{g}_{k'}}{\|\B{g}_k\|} = \sqrt{\beta_{k'}}\nu_{k'} \frac{\B{g}_k^H\bar{\B{h}}_{k'}}{\|\B{g}_k\|},
\end{align}
is the product of two independent Gaussian RVs, and following a
similar methodology used in the Rayleigh fading case, we obtain
\eqref{eq:PE3} for keyhole channels.


\end{document}